\documentclass[a4paper, traditabstract, referee]{aa}
\usepackage{aas_macros}
\usepackage[utf8]{inputenc}
\usepackage{natbib}
\usepackage{txfonts}
\usepackage{array}
\usepackage{wasysym} 
\usepackage{ifpdf} 
\usepackage{multirow}

\ifpdf                                  
 \usepackage[pdftex]{graphicx}          
 \usepackage{rotating}
 \usepackage[pdftex,
        bookmarks = true,               
        bookmarksnumbered = true,       
        pdfpagemode = UseNone,             
        pdfstartview = FitH,            
        pdfpagelayout = SinglePage,     
        backref,                        
        hyperfootnotes = true,          
        colorlinks = true,              
        urlcolor = blue,                
        linkcolor = red,                
        anchorcolor = black,            
        citecolor = blue,              
        pdfborder = {0 0 0}             
        ]{hyperref}                     
 \hypersetup{
        baseurl = {http://www.aanda.org}, 
        pdftitle = {The multiple planets transiting Kepler-9. I. Inferring stellar properties and planetary compositions}, 
        pdfsubject = {Multiple transiting planets in Kepler-9 system}, 
        pdfauthor = {Mathieu HAVEL, Tristan GUILLOT, Diana VALENCIA, Aurélien CRIDA}, 
        pdfkeywords = {extrasolar giant planets -- planet formation  -- KEPLER-9 -- transit -- multiple planet
                       -- super-earth -- KOI-377 -- planet evolution -- planet composition}, 
        pdfcreator = {LaTeX, hyperref}, 
        pdfproducer = {PDFLaTeX}, 
 }
 \DeclareGraphicsExtensions{.pdf, .png, .jpg} 
\else                                   
 \usepackage{graphicx}                  
 \usepackage{rotating}
 \usepackage[
        colorlinks = true,              
        urlcolor = blue,                
        linkcolor = red,                
        anchorcolor = black,            
        citecolor = green               
        ]{hyperref}                     
 \DeclareGraphicsExtensions{.eps, .ps, .pst}
\fi
\graphicspath{{figs/}}

\newcommand{\modif}[1]{#1}

\def\eg{\textit{e.g. }}

\def\mearth{\ensuremath{\rm M_\oplus}~}
\def\rearth{\ensuremath{\rm R_\oplus}~}

\def\teff{\ensuremath{T_{\rm eff}}~}
\def\teq{\ensuremath{T_{\rm eq}}~}

\def\aap{{\it Astron. \& Astrophys}}

\begin{document}

\title{The multiple planets transiting Kepler-9}
\subtitle{I. Inferring stellar properties and planetary compositions}
\titlerunning{Kepler-9: stellar properties and composition of the planets}
\authorrunning{Havel et al.}

\author{Mathieu Havel\inst{\ref{oca}} \and
        Tristan Guillot\inst{\ref{oca}} \and
        Diana Valencia\inst{\ref{oca}, \ref{mit}} \and
        Aurélien Crida\inst{\ref{oca}}}

\offprints{\href{mailto:mathieu.havel@oca.eu}{mathieu.havel@oca.eu}}

\institute{Universit\'e de Nice-Sophia Antipolis, CNRS UMR 6202, Observatoire de la
           C\^ote d'Azur, B.P. 4229, 06304 Nice Cedex 4, France\label{oca} \and
           Earth, Atmospheric and Planetary Sciences, MIT, 77 Massachusetts Ave, Cambridge, MA, 02139, USA \label{mit}}

\date{Revised article submitted to A\&A and accepted, updated on \today}

\abstract
{The discovery of multiple transiting planetary systems offers new
possibilities for characterising exoplanets and
understanding their formation. The Kepler-9 system contains two
Saturn-mass planets, Kepler-9b and 9c. Using evolution models of
gas giants that reproduce the sizes of known transiting planets
and accounting for all sources of uncertainties, we show that Kepler-9b
(respectively 9c) contains $45^{+17}_{-12}$\,\mearth\ (resp. $31^{+13}_{-10}$\,\mearth) of hydrogen
and helium and $35^{+10}_{-15}$\,\mearth (resp. $24^{+10}_{-12}$\,\mearth) of heavy elements. More
accurate constraints are obtained when comparing planets 9b and 9c:
the ratio of the total mass fractions of heavy elements are $Z_{\rm
  b}/Z_{\rm c}=1.02\pm 0.14$, indicating that, although the masses of the planets differ,
their global composition is very similar, an unexpected result for
formation models. Using evolution models for super-Earths, we find
that Kepler-9d must contain less than 0.1\% of its mass in hydrogen
and helium and predict a mostly rocky structure with a total mass
between 4 and 16\,\mearth.
}


\maketitle


\section{Introduction}

Although much progress has been made since the discovery of the first
transiting exoplanet, understanding their composition, evolution,
and formation has remained elusive. One longstanding problem has been
that a significant fraction of close-in exoplanets are
inflated compared to what theoretical models
predict \citep{BLM01, GS02, Baraffe2003, Guillot2006a, Burrows2007,
Guillot2008, Miller2009a}. As a consequence, the global composition
that may be derived from size and mass measurements of a given planet
is intrinsically model-dependent. This implies that, thus far,
constraints from the compositions and their consequences in terms of
planet formation models have only been grasped in a statistical way
\citep[e.g.][]{Ida2008, Mordasini09}, not from any analysis of
individual planetary systems.

The discovery of the multiple system of transiting planets around
Kepler-9 \citep{Holman10} opens a new window characterisation
of exoplanets and on understanding their formation. The system
consists of two Saturn-size planets with 19.2 and 38.9 day orbital
periods and a likely super-Earth candidate with a 1.6 day orbit
\citep{Torres2010}. The advantage of this system is that the planets
and the star share the same age within a few million years, therefore
we can constrain the composition of one giant planet much more accurately
relative to the other.  These two planets are also in a 2:1
mean motion resonance which means that their dynamical history
is strongly constrained. Altogether, this implies that a detailed
scenario of the formation and the dynamical and physical evolution of the
complete system may be obtained. In this first article, we focus on
the compositional constraints obtained for the three planets in the
system.

Given that the largest uncertainties in the parameters of transiting
planets arise from the uncertainties in the star
\citep[eg.][]{Sozzetti2007, Torres2008}, we first derive the stellar
properties (\autoref{sec:k9a}). We subsequently infer the compositions
of the two confirmed giant planets (\autoref{sec:k9bc}) and model
the possible composition for the small planet candidate, Kepler-9d in
\autoref{sec:k9d}.  We finish by discussing the implications of our results.

\section{Kepler-9a, a solar-like star \label{sec:k9a}}
\begin{figure}
\centerline{\resizebox{8.5cm}{!}{\includegraphics{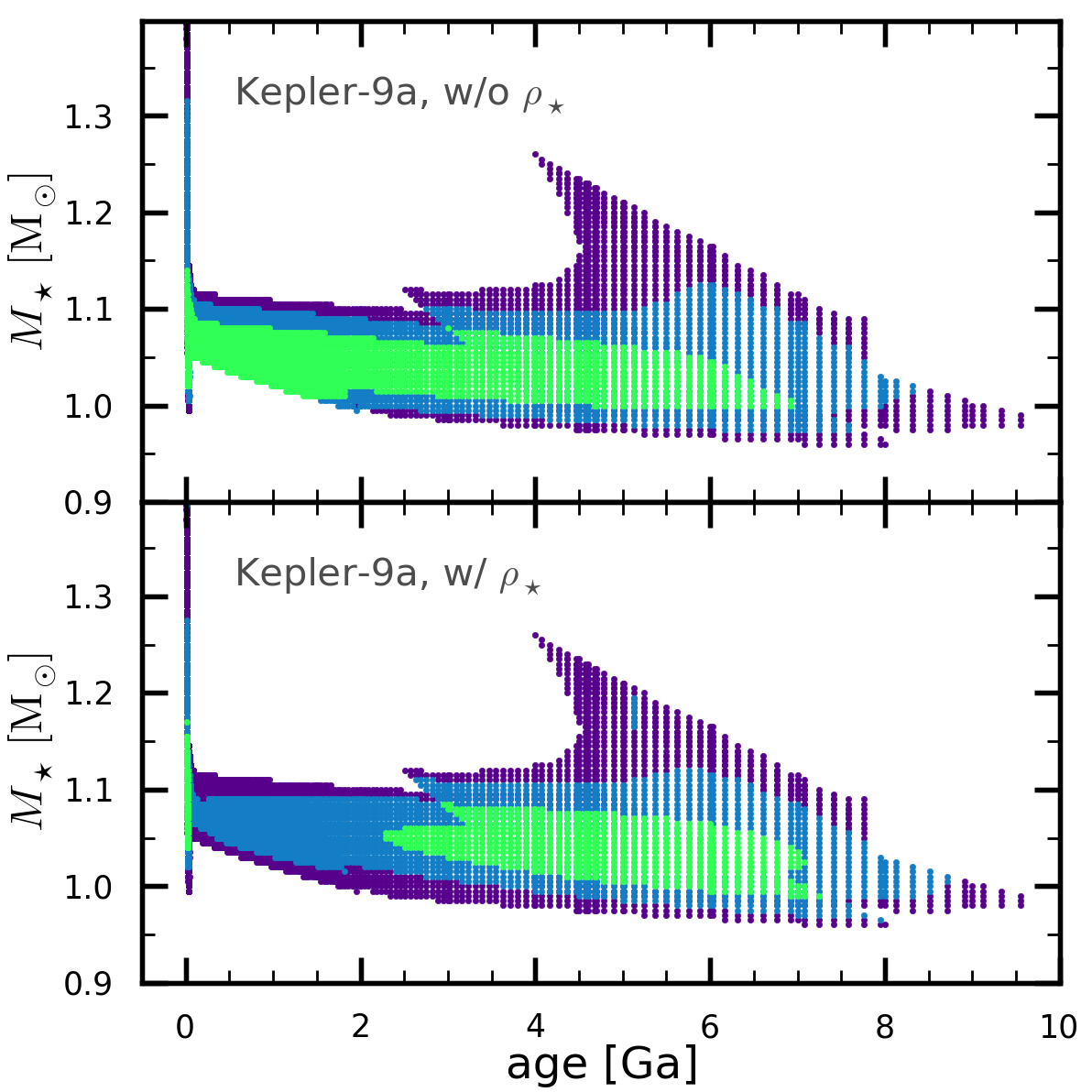}}}
\caption{Constraints derived from stellar evolution models on the mass of Kepler-9a as a function of its age.
         The dots correspond to solutions that fit the input constraints at the 68.3\%, 95.4\%
         (blue), and 99.7\% levels (purple). The upper panel uses $T_{\rm eff}$ and $\log g$
         as inputs constraints. The lower panel in addition uses the
         constraint on the stellar mean density.}
\label{fig:k9a_star}
\end{figure}

According to \citet{Holman10}, Kepler-9a, the host star of the system is a solar-like star, with an
estimated mass of $1.0\pm0.1\, \rm M_\odot$ and radius of $1.1\pm0.09\, \rm R_\odot$. Spectroscopic
measurements give a \teff of $5777\pm 61\, \rm K$ and a super-solar metallicity [Fe/H] of $0.12\pm0.04\,
\rm dex$. The star is slightly more active than our Sun, with a rotation
period of 16.7 days, implying an age of 2 to 4 Ga from gyrochronology \citep{Barnes2007, Holman10}.

We chose to re-examine the constraints on the stellar parameters with the approach described in
\citet{Guillot10Corot2}. We used the measured effective temperature and surface gravity as constraints
for the evolution models. Alone, these constraints are relatively weak
compared to what is achieved for other stars with transiting systems, because the
stellar density obtained from photometric measurement is not provided directly by
\citet{Holman10}, probably because the analysis is complex. However, it may be obtained
from the estimate of the planetary semi-major axis, orbital period, and the inferred stellar radius
\citep[eg. see][]{Beatty2007}. Our adopted final value for the density
of \modif{Kepler-9a}, $\rho_\star=0.79\pm 0.19\rho_\odot$ was obtained from the constraints provided by the two
giant planets, with the error estimated as the quadratic mean of the
two values.

Using a grid of stellar evolution models calculated with CESAM
\citep{ML08}, we determined all combinations of stellar mass, age, and
metallicity that match the constraints. We first calculated solutions
using only the constraints obtained from effective temperature and
gravity. Assuming Gaussian errors for both quantities, we derived three
ellipses corresponding to probabilities of occurrence of 68.3\%
($1\sigma$), 95.4\% ($2\sigma$), and 99.7\% ($3\sigma$),
respectively. The ensemble of solutions that fall within these values
is represented with colour-coded dots in \autoref{fig:k9a_star} (top
panel). After restricting ourselves to the ``$2\sigma$'' solutions, we see
that the stellar mass is constrained to lie within 1.0 and 1.1$\rm
M_{\odot}$, but that the age constraint is extremely weak (only ages
$>8$\,Ga are excluded). Adding the stellar density constraints
(\autoref{fig:k9a_star} bottom panel) yields a tighter constraint on
the stellar age, but very similar results in mass. \modif{The corresponding
stellar parameters of this case are summarised as a function of age 
in \autoref{tab:star}}. With the 2-4\,Ga age range obtained from gyrochronology,
we obtain a stellar mass $M_\star=1.05\pm 0.03\,\rm M_\odot$ and radius $R_\star=1.05\pm
0.06\,\rm R_\odot$, in good agreement with \citet{Holman10}.

\begin{table}[htbp]
  \begin{center}
  \begin{tabular}{c|cc|c|c}
        \hline
        \hline
        age & & $M_\star$ & $R_\star$ & $\rho_\star$ \\
        ~[Ga] & & $\rm[M_\odot]$ & $\rm[R_\odot]$ & $\rm[\rho_\odot]$ \\
        \hline
        \multirow{2}*{0.5} & 1$\sigma$ & - & - & - \\ 
                           & 2$\sigma$ & [1.05 -- 1.09] & [0.94 -- 0.99] & [1.13 -- 1.26] \\\\ 
        \multirow{2}*{1.0} & 1$\sigma$ & - & - & - \\
                           & 2$\sigma$ & [1.04 -- 1.09] & [0.94 -- 1.00] & [1.08 -- 1.24] \\\\
        \multirow{2}*{1.5} & 1$\sigma$ & - & - & - \\
                           & 2$\sigma$ & [1.02 -- 1.09] & [0.94 -- 1.02] & [1.02 -- 1.24] \\\\
        \multirow{2}*{2.0} & 1$\sigma$ & - & - & - \\
                           & 2$\sigma$ & [1.03 -- 1.09] & [0.95 -- 1.04] & [0.97 -- 1.20] \\\\
        \multirow{2}*{2.5} & 1$\sigma$ & [1.04 -- 1.06] & [0.98 -- 1.01] & [1.02 -- 1.09] \\
                           & 2$\sigma$ & [1.02 -- 1.09] & [0.95 -- 1.06] & [0.91 -- 1.18] \\\\
        \multirow{2}*{4.0} & 1$\sigma$ & [1.02 -- 1.08] & [0.99 -- 1.11] & [0.79 -- 1.04] \\
                           & 2$\sigma$ & [1.00 -- 1.11] & [0.96 -- 1.16] & [0.70 -- 1.12] \\\\
        \multirow{2}*{5.0} & 1$\sigma$ & [1.01 -- 1.07] & [1.01 -- 1.14] & [0.72 -- 0.98] \\
                           & 2$\sigma$ & [0.99 -- 1.20] & [0.97 -- 1.53] & [0.33 -- 1.07]  \\\\
        \multirow{2}*{6.0} & 1$\sigma$ & [1.00 -- 1.07] & [1.02 -- 1.18] & [0.65 -- 0.93] \\
                           & 2$\sigma$ & [0.98 -- 1.12] & [0.99 -- 1.33] & [0.48 -- 1.00]  \\\\
        \multirow{2}*{7.0} & 1$\sigma$ & [0.99 -- 1.04] & [1.05 -- 1.16] & [0.67 -- 0.86] \\
                           & 2$\sigma$ & [0.97 -- 1.09] & [1.00 -- 1.32] & [0.47 -- 0.96]  \\\\
        \multirow{2}*{8.0} & 1$\sigma$ & - & - & - \\
                           & 2$\sigma$ & [0.97 -- 1.03] & [1.03 -- 1.20] & [0.59 -- 0.89] \\\\
        \multirow{2}*{8.5} & 1$\sigma$ & - & - & - \\
                           & 2$\sigma$ & [1.00 -- 1.02] & [1.12 -- 1.19] & [0.61 -- 0.70] \\
        \hline
  \end{tabular}
  \end{center}
  \caption{Derived stellar parameters at 68.3\% and 95.4\% level of confidence
           (green and blue regions respectively, in \autoref{fig:k9a_star}).}
  \label{tab:star}
\end{table}

\section{Modelling the giant planets Kepler-9b and Kepler-9c \label{sec:k9bc}}

\subsection{Methodology}

The characteristics of the two giant planets derived by
\citet{Holman10} are $M_{\rm p} = 80.1 \pm 4.1$\,\mearth, $R_{\rm p} =
9.44 \pm 0.77$\,\rearth for Kepler-9b, and $M_{\rm p} = 54.7 \pm
4.1$\,\mearth, $R_{\rm p} = 9.22 \pm 0.75$\,\rearth for Kepler-9c.  We
derived zero-albedo equilibrium temperatures \citep[see][]{Saumon1996}
of \teq = 780 K and \teq = 620 K for the two planets, respectively
\footnote{These estimates assume circular orbit, but for $e=0.2$
time-averaged values only decrease by $\sim 0.3\%$}.

Since the semi-amplitude of the radial velocity measurements is
not provided by \citet{Holman10}, and the eccentricity of the
planet's orbits is not well constrained, we choose to not derive the
planetary mass again and use the above-mentioned values. We note that a
10-15\% change in the total mass (equivalent to $\sim 2\sigma$ of the
quoted error) of the planet induces an uncertainty on the modelled
radius of 3-4\%, a relatively significant value.
\modif{Further analysis of photometric and radial-velocity measurements should
therefore allow the derivation of tighter constraints on the planetary
mass and therefore planetary composition than possible with the data at
our disposal.}

\begin{figure}[htbp]
\centerline{\resizebox{8.5cm}{!}{\includegraphics{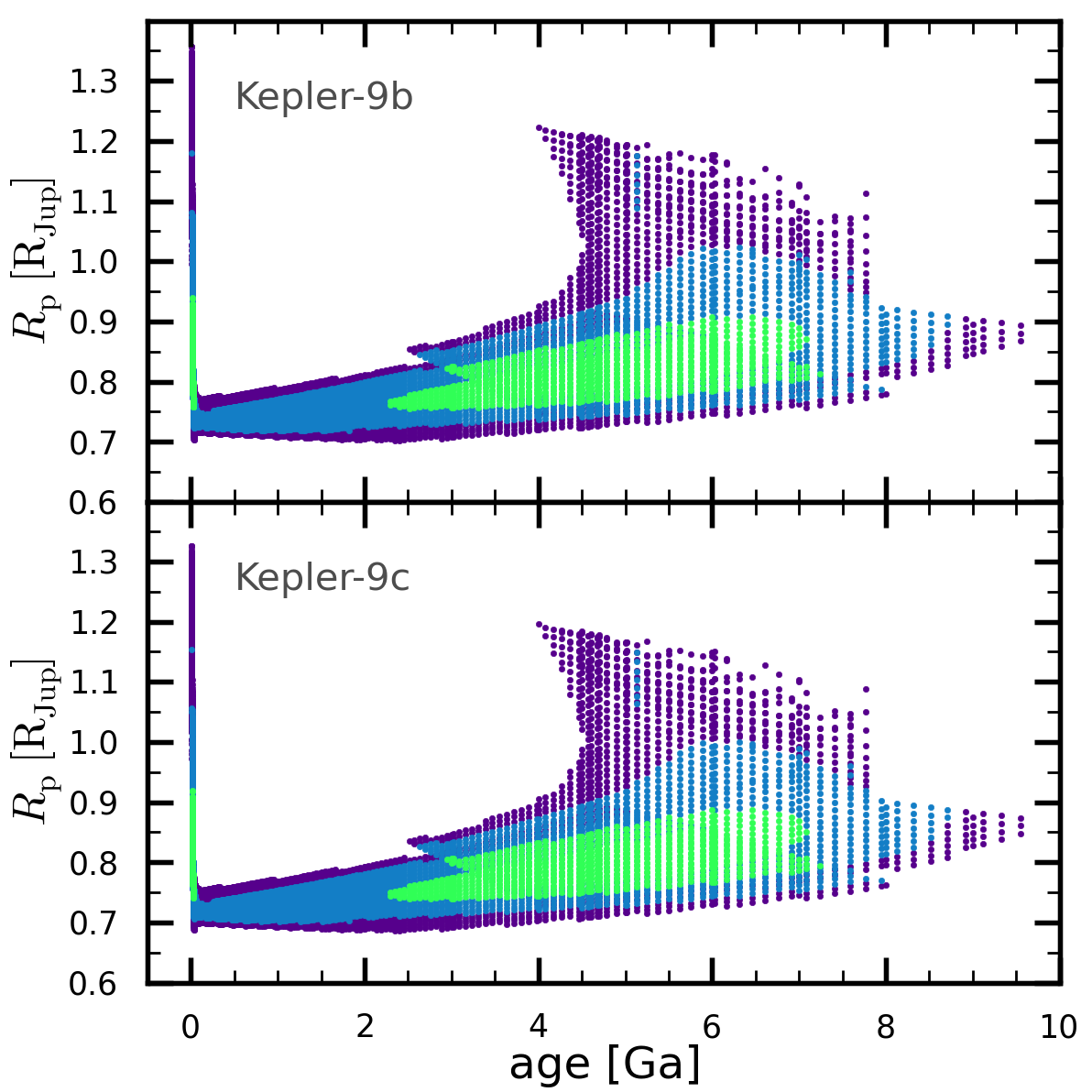}}}
\caption{Constraints derived from stellar evolution models on the radius of Kepler-9b (upper panel) and
         Kepler-9c (lower panel) as a function of its age. \modif{We use $1\rm\,R_{jup}=71,492\,km$.}
         See \autoref{fig:k9a_star} for colour coding.
         These results use solutions presented in lower panel of \autoref{fig:k9a_star}, \textit{ie.}
         with the constraint on the stellar density.}
\label{fig:k9bc_rp}
\end{figure}

On the other hand, we do use the known radii ratios provided by transit light curves
($k_{\rm b} = R_{\rm p,\ b}/R_\star = 0.07885 \pm 0.00081$ and
$k_{\rm c} = R_{\rm p,\ c}/R_\star = 0.07708 \pm 0.00080$, respectively for Kepler-9b and Kepler-9c) and
our results for the stellar radius (with the constraint on the stellar density) to compute the radius
of each planet as a function of the age, propagating all sources of
uncertainties. In the 2-4 Ga age range, we find that the two planets
have radii $R_{\rm p,\ b} =56800_{-2500}^{+3900}$\,km and $R_{\rm p,\
  c} =55800_{-2900}^{+3600}$\,km, respectively \modif{(see \autoref{fig:k9bc_rp})}.


In principle, the knowledge of both mass and size should allow a
direct determination of the planetary compositions. However, several
additional uncertainties have to be taken into account. A first
uncertainty in the modelling concerns the atmospheric boundary
condition to be used. In particular, the temperature of the deep
atmosphere that is used as a boundary condition for the interior
models depends on the greenhouse factor $\gamma^{-1}$, i.e. the ratio
between infrared and visible mean opacities that will depend on
unknown factors such as precise composition, cloud coverage and
atmospheric dynamics \citep{Hansen2008, Guillot10}. On the basis of detailed radiative
transfer calculations \citep{FLMF08, Spiegel09}, we adopt two extreme possibilities that
are scaled as a function of $T_{\rm eq}$, the zero-albedo equilibrium
temperature of the planet: either a low value $\gamma^{-1} =
1.7(\teq/2000\,{\rm K})^{-1/2}$ or a much higher one $\gamma^{-1} =
2.5(\teq/2000\,{\rm K})^{-2}$. Another significant
source of uncertainty is related to the inflation of close-in
exoplanets over what standard models predict, for which several
explanations have been put forward \modif{\citep[\eg][]{BLM01,  
GS02, Guillot2006a, Chabrier2007, Burrows2007, Guillot2008,
Laine2008, Miller2009a, Ibgui2009a, Batygin2010, Perna2010}.}
On the basis of both attempts to fit the ensemble of known transiting planets
\modif{\citep{Guillot2006a, Guillot2008}} and models for the generation and dissipation of
atmospheric kinetic energy generated by the stellar heating \modif{\citep{GS02, SG02, Batygin2010, Perna2010}},
we assume that heat is dissipated in the planet proportionally to its irradiation level.
We choose two models: either no heat is dissipated or a fraction \modif{(0.25\%) of the incoming stellar}
heat is dissipated at the centre of the planet, as required to reproduce the
sizes of known transiting exoplanets \modif{\citep{Guillot2006a, Guillot2008}.
When doing so, we assume equilibrium temperatures
are fixed, and thus do not propagate the uncertainties on the stellar parameters to
the irradiation levels of the planets, but this is clearly a weaker effect compared to
uncertainties \eg on the atmospheric models}.
\modif{Then, using the same approach as in \citet{Guillot10Corot2}, we calculate
grids of evolution models for Kepler-9b and 9c using CEPAM \citet{GM95}} for hydrogen-helium planets with various core masses for all
relevant total masses, atmospheric boundary conditions and assumptions regarding heat
dissipation. Although the core hypothesis is used for simplicity, we cannot distinguish
between heavy elements embedded in a central core or mixed throughout the envelope. The
difference is expected to be smaller than other sources of uncertainty
considered here \citep{Baraffe2008}.

The ensemble of possible compositions is obtained from a comparison
between model results and constraints on inferred planetary sizes and
ages \citep{Guillot10Corot2}: with a given set of assumptions for the
$M_{\rm b}$, $M_{\rm c}$, atmospheric model and dissipation value,
we obtain values of $M_{Z,{\rm b}}$ and $M_{Z,{\rm c}}$ matching the
age and $R_{\rm b}$, $R_{\rm c}$ values (when a solution exists).
Overall, we associate  values of $M_{Z,{\rm b}}$,
$M_{Z,{\rm c}}$, $Z_{\rm b}$, $Z_{\rm c}$, $M_{Z,{\rm b}}/M_{Z,{\rm c}}$,
$Z_{\rm b}/Z_{\rm c}$ to each solution of the stellar evolution
(given observational constraints), for the given choices of $k_{\rm b}$,
$k_{\rm c}$, $M_{\rm b}$, $M_{\rm c}$, atmospheric model, or dissipation value.

To assess the quality of the solutions, we first
identify which stellar models match the stellar constraints within 1,
2, and $3\sigma$ by assuming independent Gaussian errors for the
stellar density and effective temperature. We calculate planetary
solutions for the different extreme atmospheric boundary conditions
and dissipation rates using fiducial values for the planetary masses
and photometric $k$ values. We then account for the uncertainty on the
planetary masses and photometric $k$ values by adding models in which
these quantities have been modified by 1, 2, and $3\sigma$ from their mean
value, respectively.
Therefore, to estimate the uncertainties, we consider that $1\sigma$
solutions are obtained from the ensemble of points including $1\sigma$
stellar evolution solutions with mean values of ($k_{\rm b}$, $k_{\rm c}$,
$M_{\rm b}$, $M_{\rm c}$) and $1\sigma$ stellar evolution
solutions, which in turn ($k_{\rm b}$, $k_{\rm c}$, $M_{\rm b}$,
$M_{\rm c}$) has been changed by $\pm 1\sigma$. Solutions at $2\sigma$
(resp. $3\sigma$) are obtained from the ensemble of points including
$2\sigma$ (resp. $3\sigma$) stellar evolution solutions with mean
values of ($k_{\rm b}$, $k_{\rm c}$, $M_{\rm b}$, $M_{\rm c}$) and
$1\sigma$ stellar evolution solutions in which ($k_{\rm b}$,
$k_{\rm c}$, $M_{\rm b}$, $M_{\rm c}$) have been changed by $\pm
2\sigma$ (resp. $\pm 3\sigma$). We always consider the two possibilities
for the atmospheric models and for the dissipation value.

To assess the magnitude of the different contributions to the
global uncertainty in the solutions, we compared the range of solutions
$\Delta Y_X^{n\sigma}(\tau)$ obtained for each given age $\tau$
when only one given parameter $X$ (stellar parameters, $k_{\rm b}$, $k_{\rm c}$,
$M_{\rm b}$, $M_{\rm c}$, atmospheric model, dissipation value) is changed
by $\pm n\sigma$ to the total $n\sigma$ uncertainty $\Delta Y^{n\sigma}(\tau)$.
We note
\[
{f}_X^{n\sigma}(Y,\tau)=\Delta Y_X^{n\sigma}(\tau)/\Delta Y^{n\sigma}(\tau),
\]
where $Y$ is any of $M_{Z,{\rm b}}$, $M_{Z,{\rm c}}$, $Z_{\rm b}$,
$Z_{\rm c}$, $M_{Z,{\rm b}}/M_{Z,{\rm c}}$, $Z_{\rm b}/Z_{\rm c}$.

\subsection{Results}

The overall results are presented for representative age values and
for each of $M_{Z,{\rm b}}$, $M_{Z,{\rm c}}$, $Z_{\rm b}$, $Z_{\rm c}$,
$M_{Z,{\rm b}}/M_{Z,{\rm c}}$, $Z_{\rm b}/Z_{\rm c}$, in Tables
\ref{tab:mzb} to \ref{tab:zratios} \modif{(online material)}. The fraction (in percent)
of the uncertainty on the stellar parameters is denoted $f_\star$. The
uncertainties on both $k_{\rm b}$ and $k_{\rm c}$ have been combined
into a value $f_k$ and similarly for those on $M_{Z,{\rm b}}$ and
$M_{Z,{\rm c}}$, denoted $f_{M_{\rm tot}}$. The two atmospheric models
and two dissipation models are treated like the other errors ($f_{\rm atm}$
and $f_{\rm diss}$, respectively), but of course their contribution $\Delta
Y_X^{n\sigma}(\tau)$ is always the same regardless of n$\sigma$. The
values of $f_{\rm atm}$ and $f_{\rm diss}$ in Tables \ref{tab:mzb}-\ref{tab:zratios}
thus become progressively lower from 1$\sigma$, to 2$\sigma$, and to 3$\sigma$
solutions.

Tables \ref{tab:mzb} to \ref{tab:zratios} show that, in the 2-4 Ga age range, the solutions are
generally well behaved. When adding the $f$ values linearly, we obtain
on average 77\% of the total error. When summing them quadratically,
this mean value is 44\%. This indicates that our method probably
overestimates the errors, but we believe that the treatment is
adequate given the intrinsic difficulty in combining observational
uncertainties to model uncertainties.

One can note that, for ages beyond 4\,Ga, the solutions for the planet
parameters become less constrained. This is due both to the increased
number of solutions for the stellar parameters and to the existence of
solutions matching the planetary constraints with very low $M_Z$
values, particularly when considering the ratios $M_{Z,{\rm
    b}}/M_{Z,{\rm c}}$ and $Z_{\rm b}/Z_{\rm c}$.

\begin{figure}
\centerline{\resizebox{8.5cm}{!}{\includegraphics{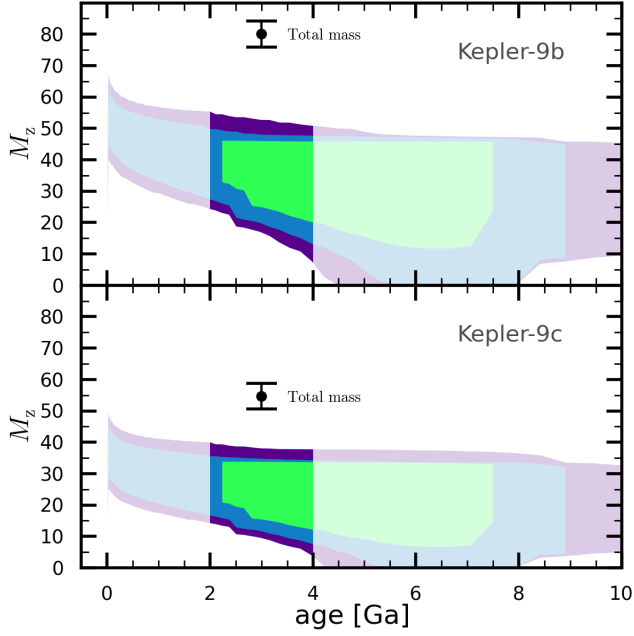}}}
\caption{Constraints derived from stellar and planetary evolution
  models on the mass of heavy elements present in planets Kepler-9b
  (top) and Kepler-9c (bottom) as a function of their age. The areas
  correspond to the ensemble of models that fit the constraints within
  $1\sigma$ (green), $2\sigma$ (blue), and $3\sigma$ (purple).  The 2-4 Ga age range is
  highlighted because it is strongly favoured by gyrochronology. The
  error bar on each panel represents the total mass of the
  corresponding planet and its $1\sigma$ uncertainty.}
\label{fig:k9bc_mcore}
\end{figure}

\autoref{fig:k9bc_mcore} shows the resulting heavy elements content in Kepler-9b and
9c as a function of age. The two planets are found to be made of
hydrogen and helium and heavy elements in relatively similar
proportions. The total masses of heavy elements needed to reproduce
the observed planetary sizes depend on the assumed age: higher masses
are required in younger systems, while pure hydrogen-helium solutions are
possible for older ages and no heat dissipation assumed. Given the 2-4 Ga age constraint,
however, the ensemble of possibilities is limited to values of $M_{Z, \rm
  b}=35^{+10}_{-15}\,\mearth$ and $M_{Z,\rm c}=
24^{+10}_{-12}\,\mearth$ when considering $1\sigma$ solutions. The constraints must be
taken with care because the ensemble of solutions has a non-Gaussian
behaviour. The values provided here overestimate slightly the ensemble
of solutions at $3\sigma$ (see Online Material for complete
solutions).

\begin{figure}
\centerline{\resizebox{8.5cm}{!}{\includegraphics{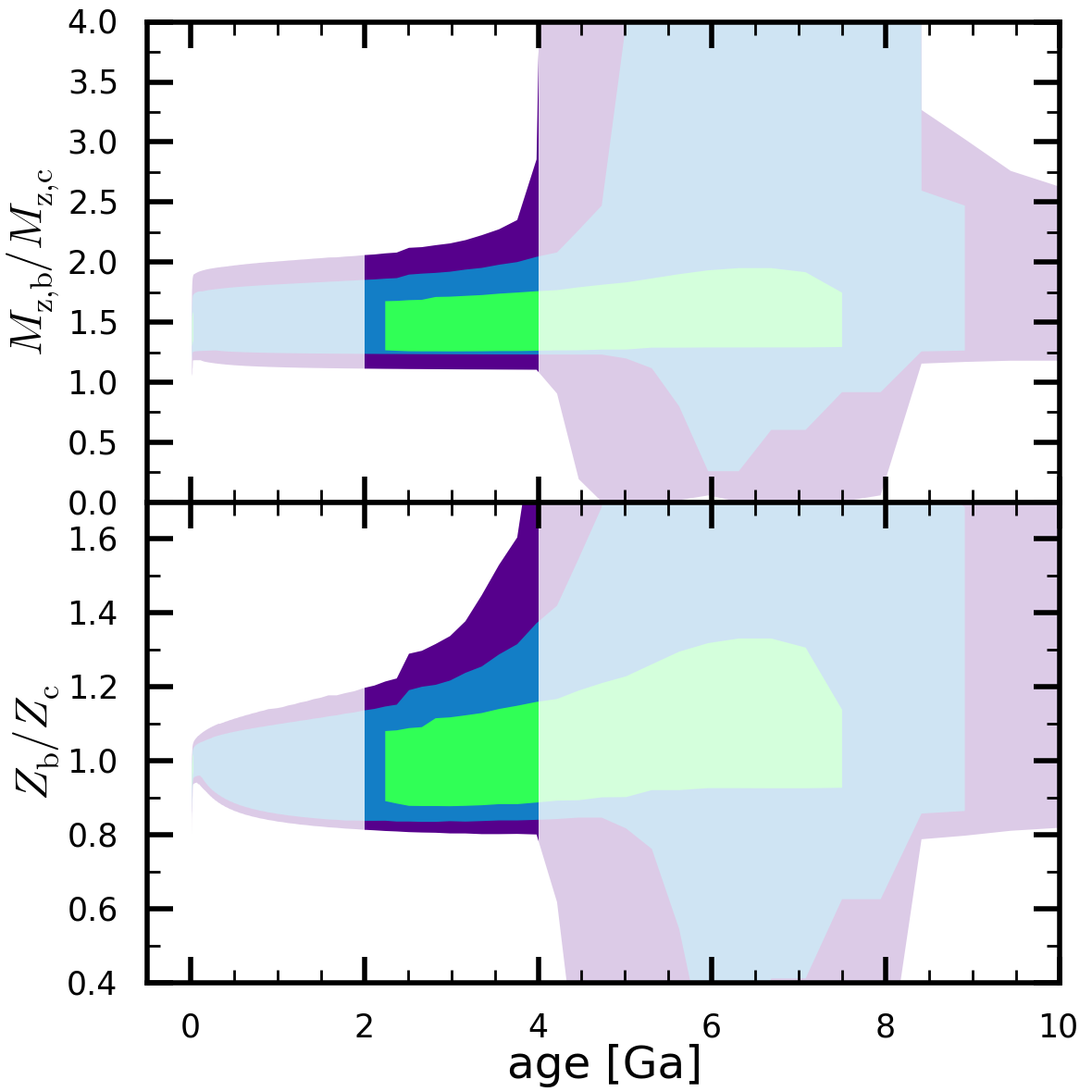}}}
\caption{Constraints on the ratios of the masses of heavy elements
  $M_{Z,{\rm b}}/M_{Z,{\rm c}}$ (top) and of the total mass fractions
  of heavy elements $Z_{\rm b}/Z_{\rm c}$ (bottom) in Kepler-9b versus
  Kepler-9c. The colours have the same meaning as in \autoref{fig:k9bc_mcore}.}
\label{fig:k9bc_ratios}
\end{figure}

When considering the planets independently, the ratio of heavy
elements to total mass of the planets are loosely bounded ($Z_{\rm
  b}/Z_{\rm c}=0.67$ to $1.81$ for $1\sigma$ solutions). However, much
tighter constraints are obtained when comparing the two planets
because the solutions are less sensitive to errors on the stellar
radius and mass. Given the similarity in mass, irradiation level, and
composition of the two planets, we also assume that the same class of
model holds and that the two planets are affected in the same way by
heat dissipation mechanisms. As shown in \autoref{fig:k9bc_ratios}, we thus find that
Kepler-9b contains $1.50\pm 0.24$ times more heavy elements in mass
than Kepler-9c. Strikingly, the two planets appear to have the same
global composition as indicated by a similar ratio of mass of heavy
elements to total planetary mass $Z_{\rm
  b}/Z_{\rm c}=1.02\pm 0.14$.

To assess the importance of the different
measurements/modelling hypotheses, we compare results obtained by
assuming only one source of uncertainty at a time to the global
results. We find that the present uncertainties on $M_Z$ and $Z$ for both
planets mostly stem from uncertainties on the stellar parameters
($\sim$50\% of the total error) and assumed dissipation rates
($\sim$30\%). When considering the ratios of these quantities,
i.e. $M_{Z,\rm b}/M_{Z,\rm c}$ and $Z_{\rm b}/Z_{\rm c}$, the contribution by the stellar
parameters' uncertainties is strongly suppressed (to less than
$\sim$10\%). In the case of $M_{Z,\rm b}/M_{Z,\rm c}$, the dominant uncertainties are
then those on the planetary masses ($\sim$60\%), on the $k$ values
($\sim$30\%), and on dissipation ($\sim$30\%). In the case of $Z_{\rm b}/Z_{\rm c}$,
the dominant uncertainties are due to the $k$ values ($\sim$40\%),
dissipation ($\sim$30\%), with less of an effect on the planetary
masses ($\sim$10\%). Further measurements from Kepler and from
ground-based radial velocimetry will be extremely valuable in reducing
the uncertainties on the stellar density, planetary masses, and $k$
values.

\section{Modelling the possible super-Earth Kepler-9d
  (KOI-377) \label{sec:k9d}}

Kepler-9d cannot be compared with the same method as its sister
planets b and c, both because of its different nature and because of
its undetermined mass. For this planet, we combined internal
structure models developed for solid planets to models of gaseous
models \citep{Valencia10}. We considered two types of planets: rocky planets with
different amounts of iron and volatile planets. For the latter we
assumed the planets to be differentiated into a solid nucleus of
terrestrial composition (a silicate mantle above an iron core),
overlaid by a gaseous envelope composed of either hydrogen and helium,
or water.  Because the equilibrium temperature we estimate for
Kepler-9d ($\teq\sim1800-2200$\,K) is well above the critical temperature of
water, the water in the envelope is in a fluid form.

\begin{figure}
\centerline{\resizebox{8.5cm}{!}{\includegraphics{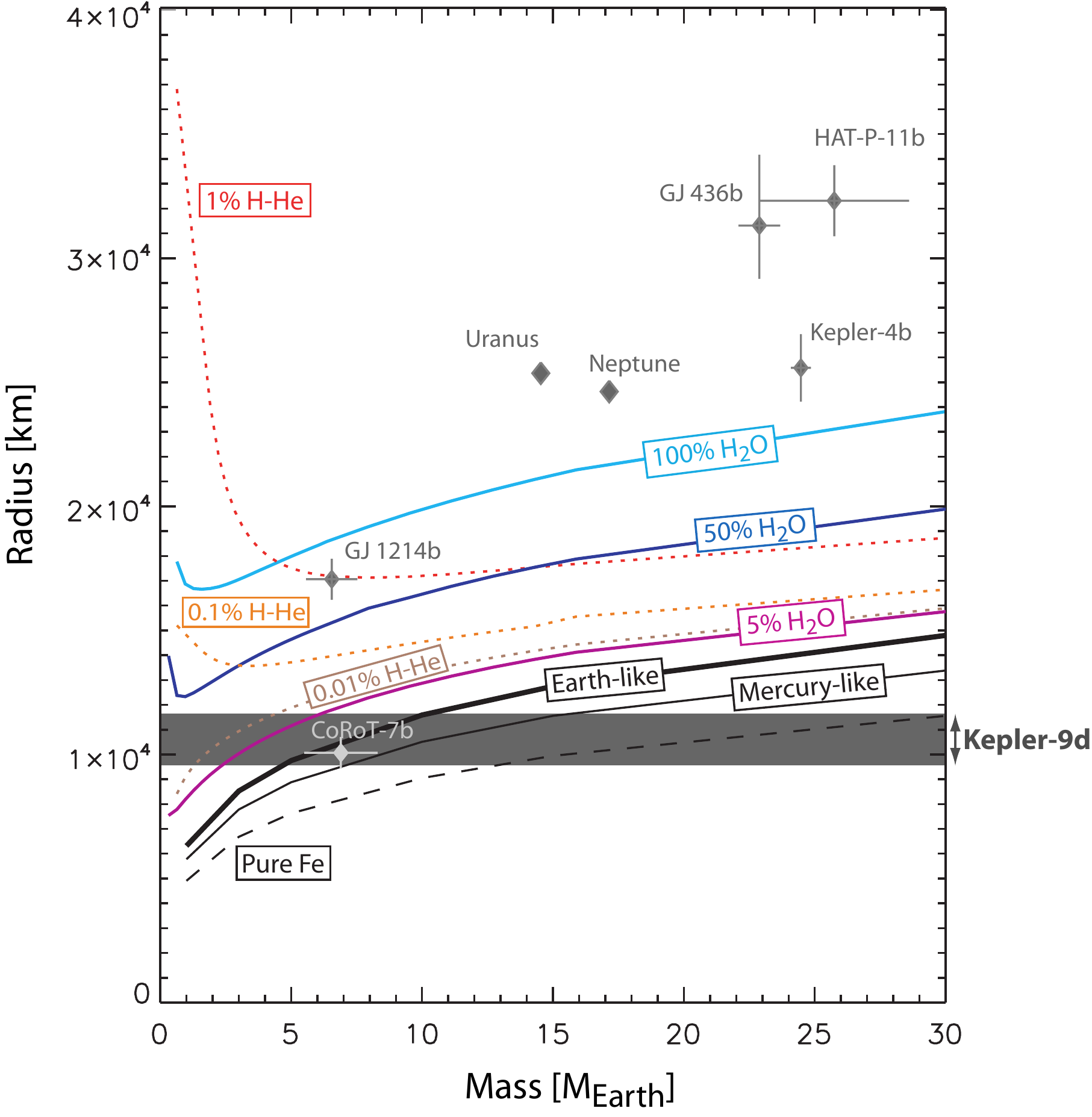}}}
\caption{Mass-radius relations for different compositions of Kepler-9d. For
the rocky scenarios: a pure iron planet (dashed black), a mercury-like
planet -- 35\% silicate mantle above a 65\% iron core (thin black),
an earth-like planet -- 63\% silicate mantle above a 33\% iron core
(thick black). For the gaseous compositions: an earth-like nucleus
covered by an H-He envelope of 0.01\% (dotted brown), 0.1\% (dotted
orange), and 1\% (dotted red) by mass; or covered by a water envelope
of 5\% (solid purple) and 50\% (solid dark blue) by mass, or a pure
100\% water-vapor planet (solid light blue). The radius range for
Kepler-9d (grey band) suggests the planet has no considerable H-He.
A physically plausible upper limit for the mass is $16\:\mathrm{M_{\oplus}}$.
Transiting exoplanets in a similar mass-radius range, as well as Uranus
and Neptune, are shown for reference. }
\label{fig:MR9d}
\end{figure}

\autoref{fig:MR9d} shows the results. Given the size of Kepler-9d, the amount of
hydrogen and helium present would be less than 0.1\% by mass. Owing to
the proximity to the star, this atmosphere would be very vulnerable to
escape, thus yielding a scenario for only hydrogen and helium that is
practically unfeasible, as obtained for CoRoT-7b \citep{Valencia10}. On the other
hand, this can be a water-vapour planet with volatiles making up less than
50\% by mass.

Alternatively, the composition may be rocky, for which the mass range
corresponding to the radius will depend on the amount of iron. The
values are $4-7\mearth$ for a planet with little or no iron (i.e. a
super-Moon), $5-11\mearth$ for a terrestrial composition (iron core is 33\%
by mass), $7-16\mearth$ for a super-Mercury composition (iron core is 65\%
by mass), and up to $30\mearth$ if made of pure iron. While it is quite
unlikely for a planet to be composed only of iron, the precise amount
is unknown. If Mercury's high iron content is used as a proxy, a
reasonable upper limit to the mass of Kepler-9d is $16\mearth$.

\section{Conclusion}

In this paper, we combined stellar and planetary evolution models to
constrain the star and planets in the Kepler-9 system in a homogeneous
way. We showed that the two Saturn-like gas giants contain a
relatively significant fraction of heavy elements in their interior
($M_{Z, \rm b}=35^{+10}_{-15}\,\mearth$ and $M_{Z,\rm
  c}=24^{+10}_{-12}\,\mearth$) and that the close-in super-Earth most
probably contains no hydrogen and helium, because the low allowed
mass fraction ($<0.01\%$) would have been rapidly blown away.

Comparing the two planets Kepler-9b and 9c led us to derive tight
constraints on the ratios of heavy elements in these planets,
i.e. $Z_{\rm b}/Z_{\rm c}=1.02\pm 0.14$. This is surprising because
accretion models predict a faster accretion of hydrogen and helium and
thus a lower $Z$ value for the first formed, most massive planet
\citep[e.g.][]{Hori2010}. These two planets are also highly interesting because they
are in a 2:1 mean motion resonance, as are several other known
exoplanetary planets. Their orbital properties generally imply an early
migration in the presence of an inner-gas disc to damp any
eccentricities \citep{Crida2008}.

Detailed studies of the formation and migration of the
entire Kepler-9 planetary system thus should shed light on the mechanisms
responsible for planetary formation.

\begin{acknowledgements}
We acknowledge the support of the {\em Programme National de Plan\'etologie}
and of {\em CNES}. Computations have been done on the Mesocentre SIGAMM
machine, hosted by the {\em Observatoire de la C\^ote d’Azur}.
\end{acknowledgements}

\bibliography{kepler9}

\clearpage
\Online

\section*{Appendix A: Planetary parameters and main uncertainties for
  Kepler-9b and 9c}

\begin{table*}[htbp]
  \begin{center}
  \begin{tabular}{l|c|rrrrrrrr}
    \hline
    \hline
    ~ age & \#$\sigma$ & $M_{\rm Z,b}$ & $\delta^+M_{\rm Z,b}$ & $\delta^-M_{\rm Z,b}$ & $f_\star$ & $f_k$ & $f_{\rm M_{tot}}$ & $f_{\rm atm}$ & $f_{\rm diss}$\\
    ~[Ga]~ & & $\rm [M_\oplus]$ & $\rm [M_\oplus]$ & $\rm [M_\oplus]$ & (\%) & (\%) & (\%) & (\%) & (\%)\\
    \hline
    1.000 & 1 &\\
          & 2 &  43.61  &    9.34  &   10.63  &    33  &   18 &    44  &   20  &    0\\
          & 3 &  43.61  &   13.23  &   14.22  &    36  &   20 &    48  &   14  &    0\\
    2.000 & 1 &\\
          & 2 &  41.19  &    8.58  &   13.86  &    44  &   18 &    36  &   18  &    6\\
          & 3 &  41.19  &   14.17  &   16.87  &    46  &   19 &    40  &   12  &    0\\
    2.239 & 1 &  40.62  &    5.44  &    7.81  &     7  &   15 &    30  &   34  &   44\\
          & 2 &  40.62  &    8.77  &   14.56  &    47  &   17 &    34  &   17  &    7\\
          & 3 &  40.62  &   13.88  &   17.57  &    48  &   19 &    38  &   12  &    0\\
    2.500 & 1 &  39.74  &    6.34  &    8.81  &    22  &   13 &    26  &   29  &   35\\
          & 2 &  39.74  &    9.28  &   17.84  &    56  &   15 &    27  &   15  &    6\\
          & 3 &  39.74  &   13.86  &   20.70  &    55  &   18 &    32  &   11  &    1\\
    3.000 & 1 &  37.67  &    8.30  &   12.79  &    43  &   10 &    17  &   21  &   28\\
          & 2 &  37.67  &   10.36  &   17.96  &    58  &   15 &    25  &   14  &    9\\
          & 3 &  37.67  &   15.05  &   20.82  &    58  &   17 &    30  &   10  &    2\\
    3.500 & 1 &  36.19  &    9.68  &   13.76  &    47  &    9 &    14  &   19  &   28\\
          & 2 &  36.19  &   11.63  &   19.63  &    61  &   13 &    21  &   13  &   11\\
          & 3 &  36.19  &   15.64  &   23.82  &    62  &   16 &    25  &    9  &    3\\
    4.000 & 1 &  35.17  &   10.67  &   15.06  &    49  &    8 &    12  &   17  &   29\\
          & 2 &  35.17  &   12.51  &   22.07  &    63  &   12 &    18  &   12  &   12\\
          & 3 &  35.17  &   15.65  &   27.93  &    66  &   15 &    20  &    9  &    4\\
    5.000 & 1 &  29.89  &   15.94  &   14.41  &    52  &    7 &     9  &   14  &   29\\
          & 2 &  29.89  &   17.53  &   25.23  &    66  &   10 &    12  &    9  &   14\\
          & 3 &  29.89  &   19.02  &   29.88  &    77  &    6 &    13  &    8  &    7\\
    6.000 & 1 &  28.19  &   17.63  &   16.26  &    48  &    6 &     7  &   13  &   35\\
          & 2 &  28.19  &   19.02  &   28.12  &    70  &    5 &     9  &    9  &   16\\
          & 3 &  28.19  &   19.48  &   28.12  &    73  &    7 &    14  &    8  &   12\\
    7.000 & 1 &  27.27  &   18.54  &   14.92  &    40  &    7 &     7  &   14  &   43\\
          & 2 &  27.27  &   19.71  &   27.14  &    65  &    5 &     8  &    9  &   21\\
          & 3 &  27.27  &   20.11  &   27.26  &    69  &    7 &    12  &    9  &   17\\
    8.000 & 1 &\\
          & 2 &  27.97  &   18.48  &   26.61  &    53  &   10 &     9  &   10  &   29\\
          & 3 &  27.97  &   19.19  &   27.11  &    63  &    8 &    13  &    9  &   23\\
    \hline
  \end{tabular}
  \end{center}
  \caption{Constraints obtained on $M_{\rm Z,b}$ as a function of age and sources of uncertainties}
  \label{tab:mzb}
\end{table*}

\begin{table*}[htbp]
  \begin{center}
  \begin{tabular}{l|c|rrrrrrrr}
    \hline
    \hline
    ~ age & \#$\sigma$ & $M_{\rm Z,c}$ & $\delta^+M_{\rm Z,c}$ & $\delta^-M_{\rm Z,c}$ & $f_\star$ & $f_k$ & $f_{\rm M_{tot}}$ & $f_{\rm atm}$ & $f_{\rm diss}$\\
    ~[Ga]~ & & $\rm [M_\oplus]$ & $\rm [M_\oplus]$ & $\rm [M_\oplus]$ & (\%) & (\%) & (\%) & (\%) & (\%)\\
    \hline
    1.000 & 1 &\\
        & 2 &   29.97 &     7.66 &    9.54 &     26 &     15 &     50 &     21 &      8\\
        & 3 &   29.97 &    11.24 &   12.43 &     28 &     16 &     54 &     14 &      3\\
    2.000 & 1 &\\
        & 2 &   28.32 &     7.21 &   11.67 &     36 &     14 &     41 &     19 &     16\\
        & 3 &   28.32 &    11.67 &   14.09 &     38 &     16 &     46 &     13 &      5\\
    2.239 & 1 &   27.91 &     5.87 &    7.14 &      5 &     10 &     30 &     31 &     46\\
        & 2 &   27.91 &     7.44 &   12.11 &     38 &     14 &     39 &     18 &     16\\
        & 3 &   27.91 &    11.43 &   14.50 &     40 &     16 &     45 &     13 &      7\\
    2.500 & 1 &   27.27 &     6.53 &    7.78 &     16 &     10 &     26 &     28 &     41\\
        & 2 &   27.27 &     7.92 &   14.18 &     47 &     13 &     32 &     16 &     15\\
        & 3 &   27.27 &    11.41 &   16.38 &     47 &     15 &     38 &     12 &      8\\
    3.000 & 1 &   25.72 &     8.04 &   10.29 &     34 &      8 &     19 &     21 &     34\\
        & 2 &   25.72 &     9.05 &   14.09 &     49 &     12 &     29 &     16 &     18\\
        & 3 &   25.72 &    12.33 &   16.23 &     50 &     15 &     36 &     12 &      9\\
    3.500 & 1 &   24.64 &     9.07 &   10.88 &     38 &      7 &     16 &     20 &     33\\
        & 2 &   24.64 &     9.84 &   15.05 &     53 &     12 &     25 &     15 &     19\\
        & 3 &   24.64 &    13.18 &   17.61 &     55 &     14 &     30 &     11 &     10\\
    4.000 & 1 &   23.91 &     9.74 &   11.74 &     40 &      7 &     14 &     18 &     33\\
        & 2 &   23.91 &    10.36 &   16.36 &     56 &     11 &     22 &     14 &     19\\
        & 3 &   23.91 &    13.83 &   20.27 &     58 &     13 &     25 &     10 &     11\\
    5.000 & 1 &   20.26 &    13.32 &   11.16 &     45 &      6 &     11 &     16 &     33\\
        & 2 &   20.26 &    13.66 &   18.43 &     61 &      9 &     15 &     11 &     20\\
        & 3 &   20.26 &    17.37 &   20.24 &     65 &      5 &     17 &      9 &     13\\
    6.000 & 1 &   19.01 &    14.44 &   12.22 &     42 &      6 &      9 &     15 &     37\\
        & 2 &   19.01 &    14.70 &   18.99 &     64 &      4 &     11 &     11 &     22\\
        & 3 &   19.01 &    18.43 &   19.01 &     61 &      6 &     16 &     10 &     17\\
    7.000 & 1 &   18.44 &    14.80 &   11.38 &     35 &      6 &      9 &     16 &     42\\
        & 2 &   18.44 &    15.15 &   18.41 &     59 &      4 &     11 &     12 &     27\\
        & 3 &   18.44 &    18.68 &   18.43 &     57 &      6 &     15 &     10 &     21\\
    8.000 & 1 &\\
        & 2 &   18.79 &    14.71 &   18.30 &     50 &      6 &     11 &     12 &     32\\
        & 3 &   18.79 &    17.67 &   18.40 &     52 &      7 &     16 &     11 &     26\\
    \hline
  \end{tabular}
  \end{center}
  \caption{Constraints obtained on $M_{\rm Z,c}$ as a function of age and sources of uncertainties}
  \label{tab:mzc}
\end{table*}

\begin{table*}[htbp]
  \begin{center}
  \begin{tabular}{l|c|rrrrrrrr}
    \hline
    \hline
    ~ age & \#$\sigma$ & $Z_{\rm b}$ & $\delta^+Z_{\rm b}$ & $\delta^-Z_{\rm b}$ & $f_\star$ & $f_k$ & $f_{\rm M_{tot}}$ & $f_{\rm atm}$ & $f_{\rm diss}$\\
    ~[Ga]~ & & & & & (\%) & (\%) & (\%) & (\%) & (\%)\\
    \hline
    1.000 & 1 &\\
        & 2 &    0.55 &     0.07 &    0.11 &     46 &     26 &      4 &     27 &      0\\
        & 3 &    0.55 &     0.10 &    0.14 &     51 &     29 &      4 &     20 &      0\\
    2.000 & 1 &\\
        & 2 &    0.52 &     0.07 &    0.16 &     54 &     22 &      3 &     22 &      7\\
        & 3 &    0.52 &     0.11 &    0.19 &     60 &     25 &      4 &     15 &      0\\
    2.239 & 1 &    0.51 &     0.06 &    0.09 &      8 &     18 &      3 &     39 &     51\\
        & 2 &    0.51 &     0.08 &    0.17 &     56 &     21 &      3 &     21 &      8\\
        & 3 &    0.51 &     0.11 &    0.20 &     61 &     24 &      4 &     15 &      1\\
    2.500 & 1 &    0.50 &     0.07 &    0.10 &     24 &     15 &      2 &     32 &     39\\
        & 2 &    0.50 &     0.08 &    0.21 &     64 &     17 &      3 &     17 &      8\\
        & 3 &    0.50 &     0.11 &    0.25 &     66 &     21 &      3 &     13 &      2\\
    3.000 & 1 &    0.47 &     0.09 &    0.15 &     46 &     10 &      2 &     22 &     29\\
        & 2 &    0.47 &     0.10 &    0.22 &     64 &     16 &      3 &     16 &     10\\
        & 3 &    0.47 &     0.13 &    0.26 &     67 &     20 &      3 &     12 &      3\\
    3.500 & 1 &    0.45 &     0.11 &    0.17 &     49 &      9 &      2 &     20 &     29\\
        & 2 &    0.45 &     0.12 &    0.25 &     65 &     14 &      2 &     14 &     11\\
        & 3 &    0.45 &     0.14 &    0.30 &     69 &     18 &      3 &     11 &      4\\
    4.000 & 1 &    0.44 &     0.12 &    0.19 &     51 &      8 &      1 &     17 &     30\\
        & 2 &    0.44 &     0.13 &    0.28 &     67 &     13 &      2 &     12 &     13\\
        & 3 &    0.44 &     0.15 &    0.35 &     71 &     16 &      2 &      9 &      5\\
    5.000 & 1 &    0.37 &     0.19 &    0.18 &     54 &      7 &      1 &     15 &     30\\
        & 2 &    0.37 &     0.19 &    0.32 &     69 &     11 &      2 &     10 &     14\\
        & 3 &    0.37 &     0.21 &    0.37 &     80 &      6 &      1 &      8 &      7\\
    6.000 & 1 &    0.35 &     0.20 &    0.20 &     51 &      6 &      1 &     14 &     36\\
        & 2 &    0.35 &     0.21 &    0.35 &     73 &      5 &      1 &      9 &     17\\
        & 3 &    0.35 &     0.23 &    0.35 &     75 &      7 &      2 &      9 &     12\\
    7.000 & 1 &    0.34 &     0.21 &    0.19 &     42 &      7 &      1 &     15 &     46\\
        & 2 &    0.34 &     0.22 &    0.34 &     68 &      5 &      1 &      9 &     22\\
        & 3 &    0.34 &     0.24 &    0.34 &     71 &      7 &      1 &      9 &     17\\
    8.000 & 1 &\\
        & 2 &    0.35 &     0.20 &    0.33 &     56 &     11 &      2 &     10 &     31\\
        & 3 &    0.35 &     0.23 &    0.34 &     64 &      8 &      2 &      9 &     23\\
    \hline
  \end{tabular}
  \end{center}
  \caption{Constraints obtained on $Z_{\rm b}$ as a function of age and sources of uncertainties}
  \label{tab:zb}
\end{table*}

\begin{table*}[htbp]
  \begin{center}
  \begin{tabular}{l|c|rrrrrrrr}
    \hline
    \hline
    ~ age & \#$\sigma$ & $Z_{\rm c}$ & $\delta^+Z_{\rm c}$ & $\delta^-Z_{\rm c}$ & $f_\star$ & $f_k$ & $f_{\rm M_{tot}}$ & $f_{\rm atm}$ & $f_{\rm diss}$\\
    ~[Ga]~ & & & & & (\%) & (\%) & (\%) & (\%) & (\%)\\
    \hline
    1.000 & 1 &\\
        & 2 &    0.55 &     0.10 &    0.13 &     37 &     21 &      5 &     29 &     11\\
        & 3 &    0.55 &     0.13 &    0.17 &     42 &     23 &      5 &     21 &      5\\
    2.000 & 1 &\\
        & 2 &    0.52 &     0.12 &    0.18 &     41 &     16 &      4 &     22 &     18\\
        & 3 &    0.52 &     0.16 &    0.21 &     48 &     20 &      5 &     16 &      7\\
    2.239 & 1 &    0.51 &     0.08 &    0.11 &      6 &     13 &      4 &     40 &     60\\
        & 2 &    0.51 &     0.13 &    0.19 &     43 &     16 &      4 &     20 &     18\\
        & 3 &    0.51 &     0.17 &    0.23 &     48 &     19 &      5 &     16 &      9\\
    2.500 & 1 &    0.50 &     0.09 &    0.12 &     19 &     12 &      3 &     33 &     49\\
        & 2 &    0.50 &     0.14 &    0.24 &     50 &     13 &      4 &     17 &     17\\
        & 3 &    0.50 &     0.18 &    0.28 &     53 &     17 &      4 &     14 &      9\\
    3.000 & 1 &    0.47 &     0.12 &    0.18 &     39 &      9 &      3 &     24 &     38\\
        & 2 &    0.47 &     0.17 &    0.24 &     51 &     13 &      3 &     16 &     18\\
        & 3 &    0.47 &     0.21 &    0.28 &     54 &     16 &      4 &     13 &     10\\
    3.500 & 1 &    0.45 &     0.14 &    0.19 &     43 &      8 &      2 &     22 &     37\\
        & 2 &    0.45 &     0.19 &    0.27 &     53 &     11 &      3 &     15 &     19\\
        & 3 &    0.45 &     0.22 &    0.32 &     57 &     15 &      4 &     11 &     10\\
    4.000 & 1 &    0.44 &     0.15 &    0.21 &     45 &      7 &      2 &     20 &     37\\
        & 2 &    0.44 &     0.20 &    0.30 &     55 &     11 &      3 &     13 &     19\\
        & 3 &    0.44 &     0.24 &    0.37 &     60 &     13 &      3 &     10 &     11\\
    5.000 & 1 &    0.37 &     0.21 &    0.20 &     48 &      7 &      2 &     18 &     35\\
        & 2 &    0.37 &     0.26 &    0.34 &     59 &      9 &      2 &     11 &     19\\
        & 3 &    0.37 &     0.30 &    0.37 &     67 &      5 &      2 &     10 &     13\\
    6.000 & 1 &    0.35 &     0.23 &    0.22 &     45 &      6 &      2 &     17 &     39\\
        & 2 &    0.35 &     0.28 &    0.35 &     62 &      4 &      1 &     11 &     22\\
        & 3 &    0.35 &     0.31 &    0.35 &     63 &      6 &      2 &     10 &     17\\
    7.000 & 1 &    0.34 &     0.24 &    0.21 &     37 &      6 &      2 &     17 &     45\\
        & 2 &    0.34 &     0.29 &    0.34 &     58 &      4 &      1 &     11 &     26\\
        & 3 &    0.34 &     0.31 &    0.34 &     60 &      6 &      2 &     11 &     22\\
    8.000 & 1 &\\
        & 2 &    0.34 &     0.27 &    0.34 &     50 &      6 &      2 &     12 &     32\\
        & 3 &    0.34 &     0.30 &    0.34 &     54 &      7 &      2 &     11 &     26\\
    \hline
  \end{tabular}
  \end{center}
  \caption{Constraints obtained on $Z_{\rm c}$ as a function of age and sources of uncertainties}
  \label{tab:zc}
\end{table*}

\begin{table*}[htbp]
  \begin{center}
  \begin{tabular}{l|c|rrrrrrrr}
    \hline
    \hline
    ~ age & \#$\sigma$ & $M_{\rm Z,b}/M_{\rm Z,c}$ & $\delta^+M_{\rm Z,b}/M_{\rm Z,c}$ & $\delta^-M_{\rm Z,b}/M_{\rm Z,c}$ & $f_\star$ & $f_k$ & $f_{\rm M_{tot}}$ & $f_{\rm atm}$ & $f_{\rm diss}$\\
    ~[Ga]~ & & & & & (\%) & (\%) & (\%) & (\%) & (\%)\\
    \hline
    1.000 & 1 &\\
        & 2 &    1.47 &     0.34 &    0.23 &      1 &     27 &     88 &      8 &     17\\
        & 3 &    1.47 &     0.53 &    0.35 &      1 &     30 &     88 &      5 &     11\\
    2.000 & 1 &\\
        & 2 &    1.47 &     0.38 &    0.24 &      3 &     33 &     86 &      9 &     18\\
        & 3 &    1.47 &     0.58 &    0.36 &      3 &     36 &     86 &      5 &     11\\
    2.239 & 1 &    1.48 &     0.20 &    0.21 &      0 &     23 &     64 &     15 &     29\\
        & 2 &    1.48 &     0.39 &    0.24 &      4 &     34 &     85 &      9 &     18\\
        & 3 &    1.48 &     0.60 &    0.36 &      3 &     38 &     86 &      5 &     11\\
    2.500 & 1 &    1.48 &     0.21 &    0.22 &      1 &     23 &     62 &     14 &     28\\
        & 2 &    1.48 &     0.42 &    0.24 &      6 &     40 &     82 &      9 &     18\\
        & 3 &    1.48 &     0.64 &    0.37 &      5 &     47 &     84 &      6 &     11\\
    3.000 & 1 &    1.49 &     0.23 &    0.23 &      6 &     26 &     59 &     14 &     27\\
        & 2 &    1.49 &     0.43 &    0.26 &      8 &     43 &     81 &      9 &     18\\
        & 3 &    1.49 &     0.67 &    0.38 &      7 &     52 &     82 &      5 &     11\\
    3.500 & 1 &    1.50 &     0.24 &    0.24 &      9 &     28 &     58 &     15 &     26\\
        & 2 &    1.50 &     0.48 &    0.26 &     11 &     50 &     78 &      9 &     17\\
        & 3 &    1.50 &     0.77 &    0.39 &     10 &     69 &     80 &      6 &     10\\
    4.000 & 1 &    1.50 &     0.26 &    0.24 &     11 &     30 &     57 &     15 &     26\\
        & 2 &    1.50 &     0.55 &    0.27 &     15 &     58 &     75 &      9 &     16\\
        & 3 &    1.50 &     2.08 &    0.41 &     10 &     86 &     42 &      3 &      5\\
    5.000 & 1 &    1.58 &     0.25 &    0.31 &     18 &     35 &     54 &     15 &     23\\
        & 2 &    1.58 &     2.34 &    0.39 &     18 &     76 &     33 &      3 &      5\\
        & 3 &    1.58 &   158.46 &    1.58 &     21 &     78 &     22 &      0 &      0\\
    6.000 & 1 &    1.60 &     0.34 &    0.31 &     25 &     42 &     51 &     15 &     18\\
        & 2 &    1.60 &   191.56 &    1.34 &      8 &     73 &     54 &      0 &      0\\
        & 3 &    1.60 &   191.56 &    1.55 &      8 &     73 &     54 &      0 &      0\\
    7.000 & 1 &    1.69 &     0.24 &    0.40 &     22 &     43 &     53 &     17 &     17\\
        & 2 &    1.69 &   110.30 &    1.08 &     10 &     14 &     88 &      0 &      0\\
        & 3 &    1.69 &   302.37 &    1.68 &     31 &     55 &     21 &      0 &      0\\
    8.000 & 1 &\\
        & 2 &    1.65 &   197.99 &    0.69 &      1 &     15 &     97 &      0 &      0\\
        & 3 &    1.65 &   198.07 &    1.46 &      9 &     79 &     89 &      0 &      0\\
    \hline
  \end{tabular}
  \end{center}
  \caption{Constraints obtained on $M_{\rm Z,b}/M_{\rm Z,c}$ as a function of age and sources of uncertainties}
  \label{tab:mzratios}
\end{table*}

\begin{table*}[htbp]
  \begin{center}
  \begin{tabular}{l|c|rrrrrrrr}
    \hline
    \hline
    ~ age & \#$\sigma$ & $Z_{\rm b}/Z_{\rm c}$ & $\delta^+Z_{\rm b}/Z_{\rm c}$ & $\delta^-Z_{\rm b}/Z_{\rm c}$ & $f_\star$ & $f_k$ & $f_{\rm M_{tot}}$ & $f_{\rm atm}$ & $f_{\rm diss}$\\
    ~[Ga]~ & & & & & (\%) & (\%) & (\%) & (\%) & (\%)\\
    \hline
    1.000 & 1 &\\
        & 2 &    1.00 &     0.10 &    0.14 &      2 &     45 &     10 &     14 &     27\\
        & 3 &    1.00 &     0.14 &    0.16 &      3 &     58 &     11 &     11 &     21\\
    2.000 & 1 &\\
        & 2 &    1.00 &     0.14 &    0.16 &      5 &     47 &     12 &     13 &     26\\
        & 3 &    1.00 &     0.20 &    0.19 &      5 &     61 &     13 &      9 &     19\\
    2.239 & 1 &    1.00 &     0.08 &    0.11 &      0 &     34 &     11 &     22 &     42\\
        & 2 &    1.00 &     0.15 &    0.16 &      6 &     48 &     12 &     12 &     25\\
        & 3 &    1.00 &     0.21 &    0.19 &      6 &     62 &     14 &      9 &     19\\
    2.500 & 1 &    1.00 &     0.09 &    0.12 &      2 &     32 &     10 &     20 &     40\\
        & 2 &    1.00 &     0.19 &    0.17 &      9 &     51 &     12 &     11 &     23\\
        & 3 &    1.00 &     0.28 &    0.20 &      8 &     68 &     13 &      8 &     16\\
    3.000 & 1 &    1.01 &     0.11 &    0.13 &      8 &     34 &     10 &     19 &     36\\
        & 2 &    1.01 &     0.21 &    0.17 &     10 &     54 &     13 &     11 &     22\\
        & 3 &    1.01 &     0.33 &    0.21 &      9 &     70 &     13 &      8 &     15\\
    3.500 & 1 &    1.01 &     0.12 &    0.13 &     11 &     36 &     10 &     19 &     34\\
        & 2 &    1.01 &     0.27 &    0.18 &     13 &     57 &     13 &     10 &     20\\
        & 3 &    1.01 &     0.50 &    0.21 &     12 &     77 &     13 &      6 &     12\\
    4.000 & 1 &    1.02 &     0.14 &    0.13 &     14 &     38 &     11 &     19 &     32\\
        & 2 &    1.02 &     0.36 &    0.18 &     16 &     61 &     13 &      9 &     17\\
        & 3 &    1.02 &     1.43 &    0.23 &     10 &     88 &      8 &      3 &      5\\
    5.000 & 1 &    1.07 &     0.15 &    0.17 &     22 &     42 &     12 &     17 &     27\\
        & 2 &    1.07 &     1.60 &    0.26 &     18 &     76 &     10 &      3 &      5\\
        & 3 &    1.07 &   108.16 &    1.07 &     21 &     78 &     15 &      0 &      0\\
    6.000 & 1 &    1.08 &     0.24 &    0.16 &     28 &     47 &     13 &     17 &     21\\
        & 2 &    1.08 &   119.62 &    0.91 &      9 &     80 &     48 &      0 &      0\\
        & 3 &    1.08 &   119.62 &    1.05 &      9 &     80 &     48 &      0 &      0\\
    7.000 & 1 &    1.14 &     0.17 &    0.22 &     25 &     48 &     14 &     19 &     19\\
        & 2 &    1.14 &    63.74 &    0.73 &     12 &     16 &     86 &      0 &      0\\
        & 3 &    1.14 &   206.39 &    1.14 &     31 &     55 &     14 &      0 &      0\\
    8.000 & 1 &\\
        & 2 &    1.11 &   114.58 &    0.46 &      2 &     18 &     97 &      0 &      0\\
        & 3 &    1.11 &   120.91 &    0.98 &     10 &     89 &     83 &      0 &      0\\
    \hline
  \end{tabular}
  \end{center}
  \caption{Constraints obtained on $Z_{\rm b}/Z_{\rm c}$ as a function of age and sources of uncertainties}
  \label{tab:zratios}
\end{table*}

\end{document}